\documentclass[prd,twocolumn,a4paper,superscriptaddress]{revtex4}
\usepackage{graphicx,amssymb}
\begin{document}
%My commands
\newcommand{\be}{\begin{equation}}
\newcommand{\ee}{\end{equation}}
\newcommand{\bq}{\begin{eqnarray}}
\newcommand{\eq}{\end{eqnarray}}
\newcommand{\bsq}{\begin{subequations}}
\newcommand{\esq}{\end{subequations}}
\newcommand{\bc}{\begin{center}}
\newcommand{\ec}{\end{center}}
\newcommand {\R}{{\mathcal R}}
\newcommand{\al}{\alpha}
\newcommand\lsim{\mathrel{\rlap{\lower4pt\hbox{\hskip1pt$\sim$}}
    \raise1pt\hbox{$<$}}}
\newcommand\gsim{\mathrel{\rlap{\lower4pt\hbox{\hskip1pt$\sim$}}
    \raise1pt\hbox{$>$}}}

\title{One-scale Model for Domain Wall Network Evolution}
\author{P. P. Avelino}
\email[Electronic address: ]{ppavelin@fc.up.pt}
\affiliation{Centro de F\'{\i}sica do Porto, Rua do Campo Alegre 687, 4169-007 Porto, Portugal}
\affiliation{Departamento de F\'{\i}sica da Faculdade de Ci\^encias
da Universidade do Porto, Rua do Campo Alegre 687, 4169-007 Porto, Portugal}
\author{C.J.A.P. Martins}
\email[Electronic address: ]{C.J.A.P.Martins@damtp.cam.ac.uk}
\affiliation{Centro de F\'{\i}sica do Porto, Rua do Campo Alegre 687, 4169-007 Porto, Portugal}
\affiliation{Department of Applied Mathematics and Theoretical 
Physics,
Centre for Mathematical Sciences,\\ University of Cambridge,
Wilberforce Road, Cambridge CB3 0WA, United Kingdom}
\author{J.C.R.E. Oliveira}
\email{jeolivei@fc.up.pt}
\affiliation{Centro de F\'{\i}sica do Porto, Rua do Campo Alegre 687, 4169-007 Porto, Portugal}
\affiliation{Departamento de F\'{\i}sica da Faculdade de Ci\^encias
da Universidade do Porto, Rua do Campo Alegre 687, 4169-007 Porto, Portugal}

\date{22 July 2005}

\begin{abstract}
We introduce a new phenomenological one-scale model for the evolution 
of domain wall networks, and test it against high-resolution field theory numerical simulations. We argue that previous numerical estimates of wall velocities are inaccurate, and suggest a more accurate method of measurement. We show that the model provides an adequate approximation to the evolution of key parameters characterizing the evolution of the network. 
We use the model to study possible scaling solutions for domain wall networks, and discuss some of their cosmological consequences.
\end{abstract}
\maketitle

\section{\label{intr}Introduction}

Topological defects \cite{Kibble,vsh} are an unavoidable consequence of
cosmological phase transitions. Understanding their formation, evolution and cosmological consequences is therefore a crucial part of any serious attempt to understand the early universe. Most studies to date have focused on cosmic strings, which at least in standard scenarios are cosmologically benign, but other elements of the defect zoo are also of interest.

A case in point is that of domain walls. Possibly the main reason for their relative neglect is that it was known almost from the outset that observational constraints rule them out if their symmetry breaking scale is $\eta\ge 1MeV$ \cite{Zeldovich}. On the other hand, later on it has been claimed that non-standard domain walls can in fact have interesting cosmological roles, in particular as a realization of the so-called 'solid universe' models \cite{Solid}. More recently, further motivation for the study of cosmic defects has emerged from fundamental theory, namely in the context of brane inflation \cite{Jones,Sarangi}. As far as is presently known, in the more realistic such scenarios it is possible to argue convincingly, using the Kibble mechanism, that only cosmic strings will form. Nevertheless, one can certainly conceive of alternative scenarios where domain walls or monopoles could also form \cite{Matsuda,Barnaby}.

Here, after a very brief introduction to domain walls (Sect. \ref{field}) we extend our recent work on domain wall networks \cite{Oliveira1,Oliveira2} by deriving (in Sect. \ref{model}) an analytic model for their evolution, in the same spirit of the model of Martins and Shellard for cosmic strings \cite{ms1a,ms1b,extend}. The large-scale features of the network are therefore characterized by a lengthscale (or correlation length) $L$ and a microscopically averaged (root-mean-squared) velocity $v$. We will then provide a thorough discussion of the possible cosmological scaling solutions of such networks (Sect. \ref{scal}) as well as testing the model by comparing it with high-resolution field theory Press-Ryden-Spergel-type \cite{Press} simulations \cite{Oliveira1,Oliveira2}. In particular, it will be shown that the standard (PRS) method of estimating wall velocities is inaccurate, and a more reliable method will be introduced. Finally (in Sect. \ref{conc}) we will revisit some key cosmological consequences of these networks, and present our conclusions.

Since the derivation of the one-scale model for domain walls will be done by analogy with that for cosmic strings and, as we shall see, their respective evolution equations are very similar, we will pursue the analogy further by discussing the domain wall and cosmic string contexts side by side at various other points along the paper. The purpose of this is twofold. First, it shows that for each physically allowed regime or scaling solution for the cosmic string case, there will be a qualitatively analogous one in the domain wall case. Hence the existing knowledge about the former case can be helpful in understanding the latter. But secondly, it will also show that despite this qualitative similarity there are differences in each case at the quantitative level. Indeed, there are cases where the different co-dimension leads to substantial physical differences, and this is why in some cases a regime that is cosmologically benign for one defect type can be cosmologically disastrous for the other.

In particular it will be seen that while for cosmic strings the linear scaling solution $L\propto t$ is an ubiquitous attractor (both in standard scenarios where strings are benign and in non-standard ones where they dominate the universe---see \cite{nonint}), for domain walls the linear scaling regime, although possible, is all but irrelevant cosmologically, and universe domination is their long-term attractor. Throughout the paper we shall use fundamental units, in which $c=\hbar=1$.

\section{\label{field}Domain wall evolution}

Domain walls arise in models with spontaneously broken discrete symmetries \cite{Kibble,vsh}. A simple example is that of a scalar field $\phi$ with the Lagrangian
\begin{equation}
\mathcal{L}={\frac{1}{2}}\phi_{,\alpha}\phi^{,\alpha}-V(\phi)\,,
\label{action1}
\end{equation}
where the potential $V(\phi)$ has a discrete set of degenerate minima, say for example
\begin{equation}
V(\phi)=V_{0}\left({\frac{\phi^{2}}{\phi_{0}^{2}}}-1\right)^{2}\,.
\label{potential}
\end{equation}
By varying the action
\begin{equation}
S=\int dt\int d^{3}x{\sqrt{-g}}\mathcal{L}\,\label{action2}
\end{equation}
with respect to $\phi$ we obtain the field equation of motion
\begin{equation}
{\frac{{\partial^{2}\phi}}{\partial 
t^{2}}}+3H{\frac{{\partial\phi}}{\partial 
t}}-\nabla^{2}\phi=-{\frac{{\partial 
V}}{\partial\phi}}\,,\label{dynamics}
\end{equation}
where $\nabla$ is the Laplacian in physical coordinates and 
$H=(da/dt)/a$ is the Hubble parameter.

In many cosmological contexts of interest, one can neglect the domain wall thickness when compared to its other dimensions, and thus treat the wall as an infinitely thin surface. With this assumption, its spacetime history can be represented by a 3D worldsheet $x^\mu=x^\mu(\zeta^a), a=0,1,2$. A new action can then be easily derived, see for example \cite{vsh}. In the vicinity of the worldsheet a convenient coordinate choice is the normal distance from the surface. Noticing that in the thin wall limit all fields in the Lagrangian should depend only on this normal coordinate, and integrating out this dependence, one finds
\begin{equation}
S=-\sigma\int d^3\zeta\sqrt{\gamma}\,,
\label{ngn}
\end{equation}
where
\begin{equation}
\gamma_{ab}=g_{\mu\nu}x^\mu_{,a}x^\nu_{,b}\,
\label{defgam}
\end{equation}
is the worldsheet metric, with the obvious definition $\gamma=det(\gamma_{ab})$, and $\sigma$ is the mass per unit area of the wall. Notice that this action is proportional to the 3-volume of the wall's worldsheet, and hence is clearly the analogue of the Goto-Nambu action for strings. Corrections to the action due to the finite width of the wall have been discussed in \cite{ghg}.

The equations of motion for a domain wall can then be derived by the well-known process of varying the action, yielding in this case
\begin{equation}
\gamma^{-1/2}\partial_a\left(\sqrt{\gamma}\gamma^{ab}x^\mu_{,a}\right)+
\Gamma^\mu_{\nu\sigma}\gamma^{ab}x^\nu_{,a}x^\sigma_{,b}=0\,.
\label{microeom}
\end{equation}
Since the wall action is invariant under worldsheet re-parametrisation, we are free to impose three arbitrary gauge conditions. However, no such conditions have been found that will lead to equations that can be readily solved, either analytically or numerically, as in the case of cosmic strings---see \cite{kawano} for an approximate analysis. As a consequence, much less is known about the dynamics of domain walls than about the dynamics of strings. In passing, however, we mention a curious point of contact between the two cases: in flat (Minkowski) spacetime, any planar string solution can be trivially turned into a domain wall solution by simply translating in the direction perpendicular to the plane---see \cite{vsh} for a discussion.

\section{\label{model}The one-scale model}

We now concentrate on analytic modelling, and start by presenting a simple phenomenological derivation of a one-scale model for defect evolution.
We will see that this easily reproduces the broad features of one-scale models for cosmic strings \cite{Kibble,ms1a,ms1b,extend} (although strictly speaking it is not self-sufficient in this regard), but has the advantage of allowing a straightforward extension to the case of domain walls. Other approaches to the analytic modelling of cosmic strings and domain walls can be found in \cite{ack,Hindmarsh}.

Let us start by considering a network of non-interacting straight infinite 
strings, oriented along a fixed direction and all with the same value of the 
velocity $v$, in a flat FRW universe. Since the momentum per unit comoving 
length is proportional to $a^{-1}$ we have $v \gamma \propto a^{-2}$ or 
equivalently
\begin{equation}
\frac{dv}{dt}+2H(1-v^2)v=0\,,
\label{vevol}
\end{equation}
where $t$ is the physical time, $a$ is the scale factor, 
$H \equiv {\dot a}/a$ is the Hubble parameter and $\gamma=(1-v^2)^{-1/2}$. On the other hand, since the average number of strings in a fixed comoving volume should be conserved, one has 
$\rho \propto \gamma a^{-2}$ or equivalently 
\begin{equation}
\frac{d \rho}{dt}+2H(1+v^2)\rho=0\,,
\label{rhoevol}
\end{equation}
where $\rho$ is the average energy density in cosmic strings and we have also 
used Eqn. (\ref{vevol}) to obtain (\ref{rhoevol}). Note that although 
we are assuming the strings to be infinite and straight we may define a 
characteristic length scale, 
\begin{equation}
L^2=\frac{\mu}{\rho},
\end{equation}
which is directly related to the average distance between adjacent strings. Here we have also defined a string mass per unit length, $\mu$.

Of course, this case with non-intersecting straight infinite 
strings all aligned along a fixed direction and having the same value of 
the velocity is completely unrealistic. In practical situations the 
special length scale $L$, defined above, will also be approximately 
equal to the curvature scale of the strings, and the strings will 
have a non-zero probability of crossing and interacting with each other. 
Also the value of the velocity will vary along the strings but in the 
absence of interactions Eqn. (\ref{rhoevol}) would remain valid, with 
$v$ being taken as the RMS velocity of the strings.

Hence, we need to add further terms to Eqns. (\ref{vevol}) and 
(\ref{rhoevol}) in the context of realistic models. Let us consider 
the latter first. The probability of a string element of size 
$L$ encountering one other segment of the same size within a time $dt$ is 
proportional to $vdt/L$ . We may therefore add a new term to the 
right hand side of (\ref{rhoevol}), which will account for the energy lost from the long string network due to loop production, so that the equation now becomes
\begin{equation}
\frac{d \rho}{dt}+2H(1+v^2)\rho=-{\tilde c}\frac{v}{L}\rho\,,
\label{rhoevol1}
\end{equation}
or, writing it in terms of the length scale $L$,
\begin{equation}
2\frac{d L}{dt}=2(1+v^2)HL+{\tilde c}v\,.
\label{stringl}
\end{equation}
This is of course under the assumption that strings do intercommute when they interact---see \cite{nonint} for a discussion of alternative scenarios.
On the other hand, in the above we assumed straight infinite strings. Realistic strings will of course be curved, and their curvature will be responsible for an acceleration term  which 
also needs to be taken into account. The velocity equation is thus corrected to
\begin{equation}
\frac{dv}{dt}=(1-v^2)\left(\frac{k}{L}-2Hv\right)\,.
\label{stringv}
\end{equation}

Notice that (\ref{stringl}-\ref{stringv}) are the evolution equations of the simplest version of the VOS model \cite{ms1b,extend}---the only things missing are the fact that the curvature correction term $k$ should be velocity-dependent \cite{extend}, and additional terms accounting for effects like friction due to particle scattering, spatial curvature and so on. It is worth stressing at this point that one of the basic assumptions of a one-scale model is, not surprisingly, that there is a single large relevant lengthscale in the problem. In addition to the characteristic lengthscale $L$ defined above (which is essentially a parametrization of the energy density in strings) one can define a correlation length $\xi$ and a curvature radius $R$, for example, and one can certainly envisage them being different---in fact this can be confirmed numerically \cite{moore}, and other approaches to modelling do allow for it \cite{ack}. However, from the point of view of one-scale modelling, one is assuming that $L=\xi=R$.

Despite the simplicity of the argument, the above derivation shows that this reasoning is robust, which is useful since we can easily apply it to the case of domain wall networks. Again we start by considering  a network of non-interacting infinite planar domain walls oriented along a fixed direction and all with the same value of the velocity $v$ in a  flat FRW universe. The momentum per unit comoving area is proportional to $a^{-1}$ so that we have $v \gamma \propto a^{-3}$ or equivalently
\begin{equation}
\frac{dv}{dt}+3H(1-v^2)v=0\,.
\label{vevoldw}
\end{equation}
As before we will assume that the average number of domain walls in a fixed 
comoving should be conserved so that 
$\rho \propto \gamma a^{-1}$ or equivalently 
\begin{equation}
\frac{d \rho}{dt}+H(1+3v^2)\rho=0\,,
\label{rhoevoldw}
\end{equation}
where $\rho$ is now the average energy density in domain walls and we have again
used Eqn. (\ref{vevoldw}) to obtain (\ref{rhoevoldw}). Similarly,
although we are assuming the domain walls to be infinite and planar we 
may define a characteristic length scale, 
\begin{equation}
L={\frac{\sigma}{\rho}}\,,
\end{equation}
which is directly related to the average distance between adjacent walls 
measured in the frame comoving with the expansion of the universe, and $\sigma$ is now the domain wall mass per unit area. Finally, making analogous modifications to allow for energy losses and acceleration due to the wall curvature, we find 
\begin{equation}
\frac{dL}{dt}=(1+3v^2)HL+c_wv\,
\label{rhoevoldw1}
\end{equation}
and
\begin{equation}
\frac{dv}{dt}=(1-v^2)\left(\frac{k_w}{L}-3Hv\right)\,;
\label{vevoldw1}
\end{equation}
the latter equation has also been previously obtained, using a different approximation,  in \cite{kawano}.

Note that there is a difference between the two cases when it comes to energy losses by intercommuting. String loops are dynamically very important due to the existence of a range of trajectories that are not self-intersecting, and hence are long-lived. No such solutions are known, or believed to exist, for domain walls, so whenever closed walls (also called `vacuum bags') are produced they will decay very quickly---a point already made by \cite{Zeldovich}.

These therefore provide a phenomenological model for domain wall evolution. As in the case of strings, we may hope that the energy loss efficiency $c_w$ may be constant and possibly independent of the cosmological epoch, but by the same token we expect the curvature parameter $k_w$ to be a velocity-dependent function. In the case of cosmic strings, one can use a combination of field theory \cite{pst,vhs2,moore} and Goto-Nambu numerical simulations \cite{bb,as} to provide a successful calibration, leading to the VOS model \cite{moore,prl}. For domain walls, however, only field theory simulations are available \cite{Press,Coulson,Larsson,Fossils,Inhomog,Garagounis,Oliveira1,Oliveira2} (though a simulation in the thin wall approximation has been attempted in \cite{kawano}), so no similarly accurate calibration will be possible. Still we will show that as far as one can ascertain, the model does provide an adequate description of our numerical simulations. Before this however, we will explore the possible scaling solutions of the model.

\section{\label{scal}Scaling solutions}

We now use the evolution equations (\ref{rhoevoldw1},\ref{vevoldw1}) of our phenomenological model to discuss the cosmological evolution of domain wall networks, in particular discussing all relevant scaling solutions. The derivation of these scaling laws is mostly analogous to what has been done for cosmic string networks in the case of the VOS model. Hence we will present each solution in turn and discuss the physical mechanisms behind them, but will not in general present detailed derivations. We refer the reader to the original papers on cosmic string networks \cite{ms1a,ms1b,extend,nonint} for more detailed discussions. Most of the derivations are fairly straightforward, even if somewhat tedious.

Let us start by neglecting the effect of the energy density in the domain walls on the background (specifically, on the Friedmann equations). As we shall shortly see this is not a good approximation, since the wall network will generally end up dominating the energy density of the universe. However it is this scenario that is effectively considered, for example, when one performs numerical simulations of domain wall networks.

In this case it is easy to see that, just as for cosmic string networks, the attractor solution to the evolution equations (\ref{rhoevoldw1},\ref{vevoldw1}) corresponds to a linear scaling solution
\begin{equation}
L=\epsilon t\,, \qquad v=const\,.
\label{defscaling}
\end{equation}
Assuming that the scale factor behaves as $a \propto t^\alpha$ the detailed form of the above linear scaling constants is
\begin{equation}
\epsilon^2=\frac{k_w(k_w+c_w)}{3 \alpha (1-\alpha)}\,
\label{scaling1}
\end{equation}
\begin{equation}
v^2=\frac{1-\alpha}{3\alpha}\frac{k_w}{k_w+c_w}\,.
\label{scaling2}
\end{equation}
As in the case of cosmic strings \cite{nonint}, an energy loss mechanism (that is, a non-zero $c_w$) may not be needed in order to have linear scaling. In fact, by considering the $c_w\to 0$ limit one finds that for
\begin{equation}
\alpha>\frac{1}{4}\,
\label{alflim}
\end{equation}
a linear scaling solution is always possible. Hence in this case a linear scaling solution may exist in both matter and radiation eras (in the case of cosmic strings this is only guaranteed to be the case in the matter era.) 
This means that having non-standard (that is non-intercommuting) domain walls is by no means sufficient to ensure a frustrated wall network.

On the other hand, if $\alpha\le 1/4$ then an energy loss mechanism is necessary to have linear scaling. Note, however, that the linear scaling solutions are physically very different for cosmic strings and domain walls. In the case of cosmic strings, in the linear scaling phase the string density is a constant fraction of the background density, whereas in the case of domain walls we have
\begin{equation}
\frac{\rho_{w}}{\rho_b}\propto t\,,
\label{wallden}
\end{equation}
so the wall density grows relative to the background density, and will eventually become dominant. This happens at a time
\begin{equation}
t_\star\sim\left(G\sigma\right)^{-1}\,.
\label{tdom}
\end{equation}
Since the domain wall mass per unit area is related to the energy scale of the phase transition, $\sigma\sim\eta^{3}$, we can also write out a given epoch as
\begin{equation}
\frac{t_\star}{t_{Pl}}\sim\left(\frac{\eta}{m_{Pl}}\right)^{-3}\,;
\label{etadom}
\end{equation}
hence walls that would become dominant around today would have been formed at a phase transition with an energy scale
\begin{equation}
\eta_0\sim100 MeV\,;
\label{domnow}
\end{equation}
notice that this is \textit{two orders of magnitude larger} than the standard Zel'dovich-Kobzarev-Okun bound \cite{Zeldovich}.
It will be seen from the discussion that follows that networks that are much heavier would have become dominant well before having reached the linear scaling regime, whereas networks that are much lighter would not yet have reached the linear regime by today. Hence the range of cosmological scenarios where the linear scaling solution is of interest is quite limited.

There is, moreover, an effect which we have neglected thus far. At early times, in addition to the damping caused by the Hubble expansion, there is a further damping term coming from friction due to particle scattering off the domain walls. Phenomenologically, it can be shown \cite{vsh} that its effect can be adequately described by a frictional force per unit area
\begin{equation}
{\bf f}=-\frac{\sigma}{\ell_f}\gamma {\bf v}\,,
\label{friction}
\end{equation}
where $v$ is the string velocity. In the above we have defined a friction length scale
\begin{equation}
\ell_f=\frac{\sigma}{N_wT^4}\propto a^4\,
\label{flenght}
\end{equation}
where $T$ is the temperature of the background and $N_w$ is the number of light particles changing their mass across the walls \cite{Kibble}. If the self-coupling of the domain wall field and its couplings to the other fields are not very small, then this expression holds at all times after the wall formation. In other cases the behaviour might be slightly different very close to the phase transition \cite{vsh}, but since we are mostly interested in the behaviour of wall networks at recent times we shall neglect this subtlety. (Also $N_w$ can be effectively zero at low temperature if the wall is described by a single field.)

Just like in the case of cosmic strings \cite{ms1a,ms1b,extend} it is then easy to modify the evolution equations of our one-scale model to account for this extra friction term. They become
\begin{equation}
\frac{dL}{dt}=HL+\frac{L}{\ell_d}v^2+c_wv\,
\label{rhoevoldw2}
\end{equation}
\begin{equation}
\frac{dv}{dt}=(1-v^2)\left(\frac{k_w}{L}-\frac{v}{\ell_d}\right)\,,
\label{vevoldw2}
\end{equation}
where we have defined a damping length scale
\begin{equation}
\frac{1}{\ell_d}=3H+\frac{1}{\ell_f}\,
\label{damplen}
\end{equation}
which includes both the effects of Hubble damping and particle scattering. If no particles scatter off the walls ($N_w=0$) then the friction length scale is infinite, and the only damping term comes from Hubble damping. Note that since $\ell_f\propto a^4$, the friction term will be dominant at early times, while the Hubble term will dominate at late times, so the late-time linear scaling solution is unchanged. However, notice that the timescale when Hubble damping dominates over friction (which is also the timescale for the walls to become relativistic) is again $t_\star$ given by Eqn (\ref{tdom}). Thus we see that domain wall networks will dominate the energy density of the universe even without ever becoming relativistic or reaching the linear scaling regime.

Just as in the case of cosmic strings \cite{ms1b,extend}, there will be two possible scaling solutions (which are necessarily transient) during the friction-dominated epoch. Also as in the case of strings \cite{nonint}, these solutions will exist regardless of whether or not the walls interact with each other (that, is, whether $c_w$ is non-zero or vanishes). If the defect-forming phase transition is such as to produce a low-density network there will be an initial period where the network will be conformally stretched. The scaling laws will therefore be
\begin{equation}
L_s\propto a\,,
\label{confl}
\end{equation}
\begin{equation}
v_s\propto \frac{\ell_f}{a}\,.
\label{confv}
\end{equation}
Notice that for domain walls this gives $v\propto a^3$, whereas for cosmic strings we would have $v\propto a^2$. Going back to the domain walls we respectively have, in the radiation and matter-dominated epochs
\begin{equation}
L_r\propto t^{1/2}\,, \quad v_r\propto t^{3/2}\,
\label{confrad}
\end{equation}
and
\begin{equation}
L_m\propto t^{2/3}\,, \quad v_m\propto t^{2}\,.
\label{confmat}
\end{equation}
We emphasize that although the network is being stretched as the scale factor, and is non-relativistic, the velocities are increasing rather fast, due to the effect of the domain wall curvature. This indicates that even in the absence of other mechanisms this regime would only be a transient. The only situation where such a stretching regime could persist would be during an inflationary phase, but in that context the much faster expansion is enough to counter the wall velocities, and in fact make them decrease. It's indeed easy to see that in the case of an exponential expansion the solution of (\ref{rhoevoldw2},\ref{vevoldw2}) has the form
\begin{equation}
L_{inf}\propto a\,
\label{infl}
\end{equation}
\begin{equation}
v_{inf}\propto a^{-1}\,.
\label{infv}
\end{equation}

Following the conformal stretching regime, or perhaps right after the formation of the network if it is formed with high enough density, there is a further transient scaling regime. The inevitability of such a regime for both strings and walls was first argued for, using simple physical arguments, in \cite{Kibble}, so we shall call this the Kibble regime. In the context of velocity-dependent models the existence of the Kibble regime can be rigorously derived. The scaling solution has precisely the same form for both types of defects
\begin{equation}
L_k\propto \left(\frac{\ell_f}{H}\right)^{1/2}\,
\label{kibl}
\end{equation}
\begin{equation}
v_k\propto \left(\ell_fH\right)^{1/2}\,,
\label{kibv}
\end{equation}
although of course the friction lengthscale will not have the same form in the two cases. For domain walls we respectively have, in the radiation and matter-dominated epochs
\begin{equation}
L_r\propto t^{3/2}\,, \quad v_r\propto t^{1/2}\,
\label{kibrad}
\end{equation}
and
\begin{equation}
L_m\propto t^{11/6}\,, \quad v_m\propto t^{5/6}\,.
\label{kibmat}
\end{equation}
Notice the differences relative to the stretching regime. Here the correlation lengths grow much faster, while the velocities grow relatively more slowly. Physically, the difference between the two regimes is one of interactions and energy losses. In the stretching regime the walls are typically quite far apart, so there is very little interaction between them---typically less than one per Hubble volume per Hubble time. In the Kibble regime, on the other hand, the walls are so close together that there is a very large number of interactions---in fact there are more than in the case of the linear scaling regime. This enhanced energy loss makes the correlation length grow quite fast. The wall velocities are still non-relativistic and growing, but because regions of the network with higher velocity than average have a larger interaction probability than slower regions (theupon leaving the network) the enhanced energy loss is also responsible for making the velocities grow more slowly than in the stretching case. Still the Kibble scaling is also a transient, which in the absence of other mechanisms will necessarily end when the network becomes relativistic.

In passing, it is again interesting to compare these scaling solutions to the ones for cosmic strings, as derived in \cite{ms1b}. We notice that wall velocities become relativistic faster than those of strings, and also that wall densities grow faster relative to the background. In fact in both the stretching and Kibble regimes the wall density grows relative to the background. On the other hand, the string density grows relative to the background in the stretching regime, but \textit{decreases} relative to the background in the Kibble regime. Thus the friction-dominated epoch will last comparatively less for domain walls than for strings.

Even allowing for friction, linear scaling would be an attractor of the above equations if one neglected the effect of the wall density on the expansion of the universe. However, we have seen that in every scaling regime considered the wall density grows relative to the background, so that a wall density term
\begin{equation}
\rho_w=\frac{\sigma}{L}\,
\label{wallden2}
\end{equation}
must be included in the Einstein equations. This changes the situation for it is easy to see that the domain wall network will eventually dominate the energy density of the universe (unless some mechanism like a subsequent phase transition were to make it decay and disappear).

Thus we again see that linear scaling is of little practical importance, since it is never reached for any cosmologically realistic network. Heavy domain walls will quickly dominate the energy density of the universe, thus changing the behaviour of the Friedmann equation, typically before linear scaling is reached---in any case any networks where it could have been reached already must be sufficiently heavy to be observationally ruled out. Light walls, on the other hand, won't yet have reached that solution: their dynamics will still be friction-dominated today if they are to be cosmologically viable.

Since a domain wall network will eventually dominate the energy density 
of the universe it is important to study the dynamics of the universe in 
this regime. The expectation \cite{Zeldovich} is that the domain wall network will again become frozen in comoving coordinates with
\begin{equation}
L\propto a\,
\label{dom1}
\end{equation}
so that the scale factor should now grow as
\begin{equation}
a\propto t^{2}\,.
\label{dom3}
\end{equation}
In this case the average distance between the walls also grows as $t^{2}$ and rapidly becomes greater than the horizon. This will happen at a time that is again given by $t_\star$ in Eqn. (\ref{tdom}). An inertial observer will see domain walls moving away towards the horizon, and as walls fade away the spacetime around the observer will asymptotically approach Minkowski space. 
Notice that this solution does not depend on $c_w$---it is valid whether or not the domain walls interact.  In the case of cosmic strings it has been shown \cite{extend,nonint} that the asymptotic solution is always $L_s\propto t$ (again, whether or not the strings intercommute). The onset of domain wall domination will be studied in more detail (both analytically and numerically) in future work.

\section{\label{num}Numerical simulations}

We now proceed to compare the predictions of our analytic phenomenological model, whose evolution equations are (\ref{rhoevoldw1},\ref{vevoldw1}), with the results of numerical simulations. We shall use our own set of high-resolution field theory  numerical simulations, done on the COSMOS supercomputer and using the PRS algorithm \cite{Press}, which were first discussed in \cite{Oliveira1,Oliveira2}. As we said above, there is no analogue of the Goto-Nambu 1D simulations for the case of domain walls, though an attempt in that direction has been performed (with moderate success) in \cite{kawano}.

The fact that linear scaling is of almost no relevance in realistic cosmological scenarios involving domain walls counts against producing a well-calibrated model here, for the scenarios that one can easily simulate numerically are not the most interesting ones in practice, and vice-versa. A further difficulty with the calibration of the code using simulations with the PRS algorithm \cite{Press} is that this modifies the wall thickness. This has the side effect of erasing small-scale structures on the walls, and hence also destroying small closed walls when the wall thickness becomes comparable to their size.

\subsection{Measuring wall velocities}

A crucial point, which has been somewhat neglected in previous analyses, is that estimating the velocities of domain walls is notoriously difficult in field theory simulations---see \cite{moore} for a detailed discussion of the problems involved in the case of cosmic strings. Earlier field theory simulations, using the PRS algorithm \cite{Press} typically find
\begin{equation}
v_{prs}\sim0.4\,,
\label{vprs}
\end{equation}
no significant difference being found (when allowance is made for the magnitude of the numerical errors involved) in the values for 2D and 3D simulations.
On the other hand, Kawano \cite{kawano}, which as we have said uses a thin wall approximation (akin to the Goto-Nambu simulations for strings), and further simplifies the problem by considering only 2D simulations, typically finds smaller velocities
\begin{equation}
v_{kaw}\sim0.25-0.30\,.
\label{vkaw}
\end{equation}
We note that also in the case of cosmic strings it has been observed that velocities measured from field theory simulations tend to be somewhat higher--- again see \cite{moore} for a detailed discussion. Both types of simulations find some evidence for an approach to linear scaling, both in the radiation and matter eras. A more detailed discussing of the approach to linear scaling in domain wall field theory simulations can be found in \cite{Oliveira1,Oliveira2}.
These findings prompt us to study the issue of velocity estimations in more detail. 

The frictionless evolution of a domain wall network in a flat homogeneous and isotropic Friedmann-Robertson-Walker (FRW) universe is given by 
Eqn. (\ref{dynamics}), with the potential given by Eqn. (\ref{potential}).
If we neglet the damping term then  the solution for a plane wall with velocity $v$ is given by 
\begin{equation}
\phi=\phi_{0}\tanh\left[\frac{\pi\gamma_{0}}{\omega}
\left(z-z_{0}-v t\right)\right]\,,\label{eq:ana1}\end{equation}
where $z$ is a physical coordinate and
$\omega={\pi\phi_{0}}/{\left(2V_{0}\right)^{1/2}}$ is the
constant wall thickness in physical coordinates.
It is straightforward to show that the 
ratio between the kinetic part of the wall energy density and 
the total energy density $\rho(z)$ is independent of the $z$ coordinate 
and is equal to
\begin{equation}
v^{2}=\frac{{\dot{\phi}}^2(z)}{\rho(z)}=
\frac{\dot{\phi}^{2}}{2\gamma_{0}^{2}V(\phi)}\,,
\end{equation}
where
\begin{equation}
\rho(z)=\frac{{\dot{\phi}}^2}{2}+\frac{\left|\bigtriangledown\phi\right|^{2}}{2}+V(\phi)\,.\end{equation}

The domain walls in realistic network simulations will obviously not be planar and the velocities will vary along the walls. However, an estimate of the 
microscopic rms velocity in field theory simulations can be made using
\begin{equation}
\langle v^{2} \gamma^2 \rangle = \sum \frac{\dot{\phi}^{2}}{2V(\phi)},
\end{equation}
where the sum is limited to the points on the grid which intersect a 
domain wall. We define the wall region to be the region for which 
$-\epsilon \phi_0 <\phi < \epsilon \phi_0$ but of course there is some 
ambiguity on the choice of $\epsilon$. However, taking into account the 
result obtained for the planar wall one expects that the results will 
converge for small enough $\epsilon$ (note that we cannot realisticaly 
make $\epsilon$ arbitrarily small, since in that case we would have very poor statistics). This 
is indeed what we find. Our approximation almost completely eliminates the 
radiated energy from the walls which otherwise would contaminate 
the estimate of the velocities. This is an important advantage over 
previous estimations of $v$ where the energy radiated from the walls 
was not eliminated in the velocity estimations. We therefore expect that these previous estimates have overestimated the wall velocities.

Following \cite{Press} we have modified the equations of motion in
such a way that the walls' co-moving thickness is fixed in co-moving
coordinates in order to be able to resolve the domain walls through the 
network evolution. Ignoring the damping term due to the expansion of 
the universe the modified equations become 
\begin{equation}
\frac{\partial^{2}\phi}{\partial\eta^{2}}-\bigtriangledown_q^{2}\phi=-\frac{dV}{d\phi}\,,\label{eq:movcomovel}\end{equation}
where $\bigtriangledown_q^{2}$ is the Laplacian in comoving coordinates. The planar wall solution can now be written as 
\begin{equation}
\phi=\phi_{0}\tanh\left[\frac{\pi\gamma}{\omega}\left(q_z-q_{z0}-v\eta\right)\right]\,.\end{equation}
Notice that here $\omega$ is a fixed comoving 
thickness so that the physical wall thickness decreases with time.
Consequently, with this approximations the estimation of $\langle v^{2} \gamma^2 \rangle$ in 
realistic domain wall network simulations must now be made as 
\begin{equation}
\langle v^{2} \gamma^2 \rangle = \sum \frac{({\partial \phi}/{\partial \eta})^{2}}{2V(\phi)},
\end{equation}
rather than
\begin{equation}
\langle v^{2} \gamma^2 \rangle=\sum {\left(d\phi/dt\right)^{2}}/{V(\phi)}\,.
\end{equation}

It is easy to show from Eqn. (\ref{dynamics}) that for a planar wall solution
\begin{equation}
\frac{d{\langle {\dot \phi} \rangle}}{dt} + 3 \frac{\dot a}{a} \langle {\dot \phi} \rangle = 0\,,
\end{equation}
so that $\langle {\dot \phi} \rangle \propto a^{-3}$ and $\gamma v \propto a^{-3}$. In order 
for the momentum conservation law of the wall evolution in an expanding universe to be 
maintained one must add a damping term to Eqn. (\ref{eq:movcomovel}) so that now 
$\langle {\partial \phi}/{\partial \eta} \rangle \propto a^{-3}$ and we have
\begin{equation}
\frac{\partial^{2}\phi}{\partial\eta^{2}}+\frac{3}{a}\frac{da}{d\eta}\frac{\partial\phi}{\partial\eta} -\bigtriangledown_q^{2}\phi=-\frac{dV}{d\phi}\,.
\label{eq:movcomovel1}\end{equation}

\begin{figure}
\includegraphics[width=3.5in,keepaspectratio]{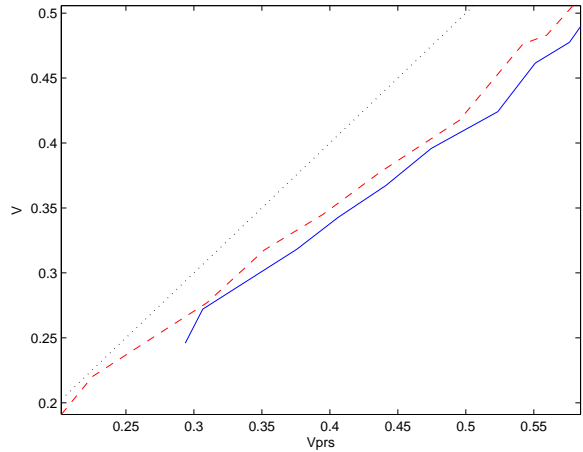}
\caption{\label{figure1}Comparing the standard (PRS) and our (new) velocity estimation methods, for 2D (solid) and 3D (dashed) radiation era runs. Each curve corresponds to an average of 10 simulations. Clearly the PRS method overestimates velocities by about 20 percent.}
\end{figure}

\begin{figure}
\includegraphics[width=3.5in,keepaspectratio]{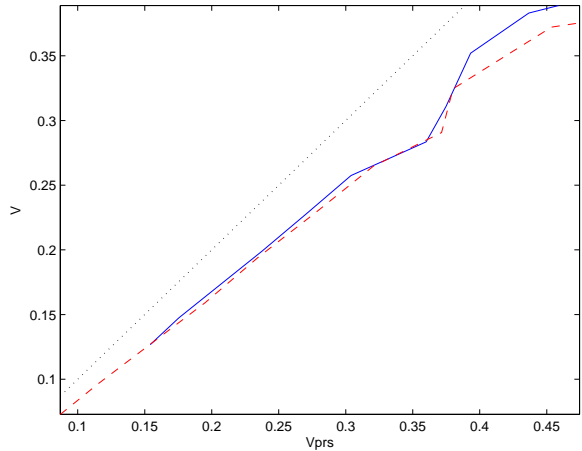}
\caption{\label{figure2}Same as Fig. \protect\ref{figure1} for matter era simulations.}
\end{figure}

The standard (PRS) method of estimating velocities can be readily compared with ours. We have thus estimated the velocities by both methods in series of 2D and 3D, radiation and matter era runs. Velocities were measured at all timesteps, and then sorted in increasing order, so that the two velocities can be ploted directly against each other without reference to time. A standard cloud-in-cell algorithm was used for data smoothing. The results of this analysis are shown in Fig. \ref{figure1} and \ref{figure2}, respectively for the radiation and matter eras.

We can see that the standard PRS method of velocity estimation typically overestimates velocities by something like 20 percent. Note that in either case the result is the same for 2D and 3D simulations (when allowance is made for numerical error bars, which were not plotted), and there is also very little difference between the radiation and matter cases. Indeed, if one fits a linear function to the above data, we find for the radiation era
\begin{equation}
v_{rad}=0.79\,v_{prs}+0.03\,,
\label{vfitsr}
\end{equation}
while in the matter era
\begin{equation}
v_{mat}=0.83\,v_{prs}+0.00\,.
\label{vfitsm}
\end{equation}
Notice that these fitting functions should not be used outside the specified range. In particular, one expects that there will be deviations from this simple linear behaviour for very small (non-relativistic) velocities.

\subsection{Testing the analytic model}

\begin{figure}
\includegraphics[width=3.5in,keepaspectratio]{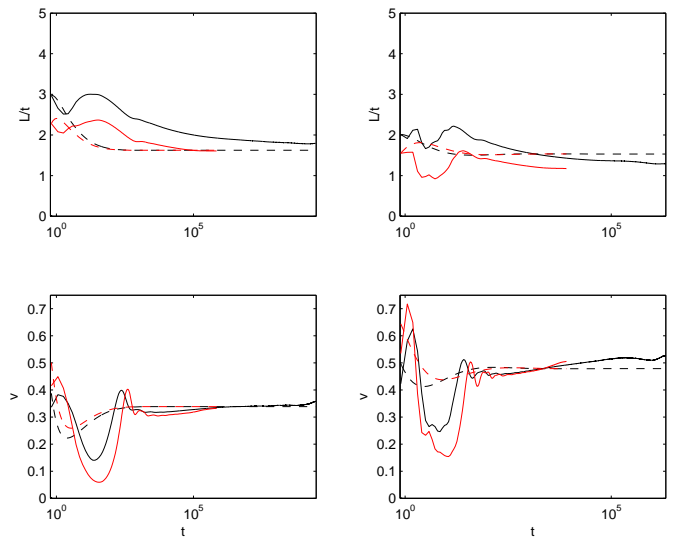}
\caption{\label{figure3}Comparing our analytic model with field theory numerical simulations in the matter (left-side panels) and radiation (right-side panels) epochs. Top panels show the evolution of the correlation length (in fact $L/t$) whereas bottom panels show the RMS velocity. In both cases the longer solid curves correspond to 2D simulations, while the shorter solid curves correspond to 3D simulations. Each such curve is in fact an averaged curve for the outcome of 100 such simulations. The corresponding dashed curves are the outcome of our analytic model, for the initial conditions relevant to each case. For all such curves we have used $c_w=0.5$ and $k_w=0.9$, which provides the best fit.}
\end{figure}

Mindful of these caveats, we have compared our analytic model with the numerical simulations of \cite{Oliveira1,Oliveira2}. We took four series of high-resolution field theory simulations, in 2D and 3D and for the radiation and matter dominated epochs. Each such series is composed of 100 different runs. For each series we have averaged the result of the 100 runs, in particular calculating averaged correlation lengths and RMS velocities that can be readily compared with the predictions of the model. As we have pointed out, although we expect the parameter $k_w$ to have some dependence on velocity, it is not easy to determine it. Hence, as a first approximation, we shall start by assuming that it is a constant, just like $c_w$. In these circumstances, we find that the best fit to the simulations is the one shown in Fig. \ref{figure3}, which corresponds to the parameters
\begin{equation}
c_w\sim0.5\,,\quad k_w\sim0.9\,.
\label{bestfit1}
\end{equation}

Several comments are in order here. The first one is that the fit is quite good (especially for the velocities), given that as we said we know that a constant $k_w$ is only an approximation. Importantly, the same parameters provide a good fit both in the radiation and in the matter epochs. (Looking for separate best fits in the two epochs would lead to only slightly different parameters, but given the inherent error bars they would effectively be indistinguishable.) Interestingly, the fit would be much worse if we had used the velocities measured by the standard PRS method.

Obviously the fit is only good at late times. At early times, one is still very close to the phase transition and fields are relaxing, so the domain wall network is not yet well-defined, and obviously the analytic model cannot be expected to account for such complicated dynamics. We could nevertheless improve the early-time fit if we had allowed for friction and radiation terms in the model---this has been successfully done for field theory simulations of cosmic string networks, as described in \cite{moore}. Given the simplicity of the present model and the approximations being made elsewhere, we think this would be an unnecessarily complication at this stage, though it should certainly be addressed in the future.

\begin{figure}
\includegraphics[width=3.5in,keepaspectratio]{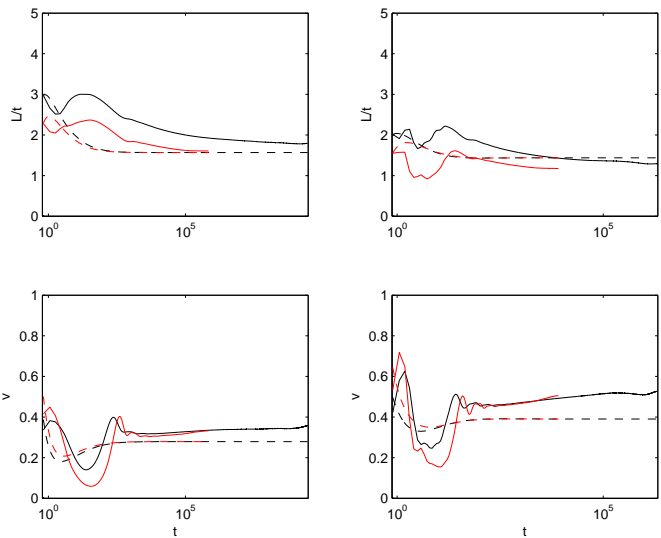}
\caption{\label{figure4}An alternative fit to the model, now using an ansatz for $k_w(v)$ that has been derived for cosmic strings---see main text for further discussion. Panel and line conventions are as in Fig. \ref{figure2}.}
\end{figure}

We emphasize that we are dealing with a phenomenological model (in fact, rather more so than in the case of cosmic strings), so one should not attach too deep a meaning to the parameters one finds. Presumably these will change if one has a proper ansatz for $k_w$. Even though we have not attempted to calculate this, we can for the sake of the argument see what happens if we assume that the function $k(v)$ that was obtained for cosmic string networks is also applicable for domain walls. This has the following form
\begin{equation}
k(v)=\frac{2\sqrt{2}}{\pi}(1-v^2)(1+2\sqrt{2}v^3)\frac{1-8v^6}{1+8v^6}\,;
\label{kstrings}
\end{equation}
we refer the reader to \cite{extend} for a detailed derivation and further discussion. Notice that with this ansatz $k$ is no longer a free parameter, and it is always smaller than unity. The result of this alternative fit is shown in Fig. \ref{figure4}, and we now find a best-fit parameter
\begin{equation}
c_w\sim1.0\,.
\label{bestfit2}
\end{equation}

As expected, this does not provide a very good fit. The energy loss term is now much larger, which has an effect on the velocities. Again a single values provides a reasonable fit for the correlation lengths in the radiation and matter epochs, but the new fit significantly underestimates the velocities. The velocity fit could of course be improved, at the expense of a poor fit for the correnation lengths.

\section{\label{conc}Conclusions}

We have introduced a new phenomenological one-scale model for the evolution 
of domain wall networks, by analogy with the stardard analytic model for cosmic string evolution. This model has then been tested against high-resolution field theory numerical simulations (in radiation and matter dominated epochs), and a good agreement has been found. Importantly, we have argued that previous numerical estimates of wall velocities are inaccurate, and tend to overestimate the wall velocities due to the effect of the radiation background. We have quantified the inherent inaccuracy (which is found to be at around the twenty percent level) and provided a more accurate method of measurement.Our phenomenological model provides an adequate approximation to the evolution of key parameters characterizing the evolution of the network, but more accurate modelling can lead to even better fitting.

We have also used the analytic model to exhaustively study possible scaling solutions for domain wall networks, and discuss some of their cosmological consequences. Comparison with analogous results for cosmic string networks has led to the identification of a number of similarities and differences between the two cases. Indeed, for each physically allowed regime or scaling solution for the cosmic string case, there is be a qualitatively analogous one in the domain wall case. However, beyond this qualitative similarity there are differences in each case at the quantitative level. The different co-dimension leads to substantial physical differences, and this is why in some cases a regime that is cosmologically benign for one defect type can be cosmologically disastrous for the other.

A case in point is that while for cosmic strings the linear scaling solution $L\propto t$ is an ubiquitous attractor (both in standard scenarios where strings are benign and in non-standard ones where they dominate the universe), for domain walls the linear scaling regime, although possible, is all but irrelevant cosmologically, and universe domination (with $L\propto t^2$) is their long-term attractor. 

Finally, let us end by noting that in the present work we have restricted ourselves to numerical simulations of domain walls in universes dominated by radiation and matter. These are of course the easiest scenarios to simulate, but plainly they are not the most relevant cosmologically. It would therefore be interesting to carry out analogous simulations of domain-wall dominated universes, not unly to understand how the domination sets in but also to test our phenomenological model in this context. We will tackle this issue in a future publication.

\section{acknowledgments}
This work was funded by FCT (Portugal), in the framework of the POCI2010 program, supported by FEDER. Specific funding came from grant POCTI/CTE-AST/60808/2004 and from the Ph.D. grant SFRH/BD/4568/2001 (J.O.)

This work was done in the context of the ESF COSLAB network. Numerical simulations were performed on COSMOS, the Altix3700 owned by the UK Computational Cosmology Consortium, supported by SGI, Intel, HEFCE and PPARC. 

\bibliography{model}

\begin{thebibliography}{30}
\expandafter\ifx\csname natexlab\endcsname\relax\def\natexlab#1{#1}\fi
\expandafter\ifx\csname bibnamefont\endcsname\relax
  \def\bibnamefont#1{#1}\fi
\expandafter\ifx\csname bibfnamefont\endcsname\relax
  \def\bibfnamefont#1{#1}\fi
\expandafter\ifx\csname citenamefont\endcsname\relax
  \def\citenamefont#1{#1}\fi
\expandafter\ifx\csname url\endcsname\relax
  \def\url#1{\texttt{#1}}\fi
\expandafter\ifx\csname urlprefix\endcsname\relax\def\urlprefix{URL }\fi
\providecommand{\bibinfo}[2]{#2}
\providecommand{\eprint}[2][]{\url{#2}}

\bibitem[{\citenamefont{Kibble}(1976)}]{Kibble}
\bibinfo{author}{\bibfnamefont{T.~W.~B.} \bibnamefont{Kibble}},
  \bibinfo{journal}{J. Phys.} \textbf{\bibinfo{volume}{A9}},
  \bibinfo{pages}{1387} (\bibinfo{year}{1976}).

\bibitem[{\citenamefont{Vilenkin and Shellard}(1994)}]{vsh}
\bibinfo{author}{\bibfnamefont{A.}~\bibnamefont{Vilenkin}} \bibnamefont{and}
  \bibinfo{author}{\bibfnamefont{E.~P.~S.} \bibnamefont{Shellard}},
  \emph{\bibinfo{title}{Cosmic Strings and other Topological Defects}}
  (\bibinfo{publisher}{Cambridge University Press},
  \bibinfo{address}{Cambridge, U.K.}, \bibinfo{year}{1994}).

\bibitem[{\citenamefont{Zeldovich et~al.}(1974)\citenamefont{Zeldovich,
  Kobzarev, and Okun}}]{Zeldovich}
\bibinfo{author}{\bibfnamefont{Y.~B.} \bibnamefont{Zeldovich}},
  \bibinfo{author}{\bibfnamefont{I.~Y.} \bibnamefont{Kobzarev}},
  \bibnamefont{and} \bibinfo{author}{\bibfnamefont{L.~B.} \bibnamefont{Okun}},
  \bibinfo{journal}{Zh. Eksp. Teor. Fiz.} \textbf{\bibinfo{volume}{67}},
  \bibinfo{pages}{3} (\bibinfo{year}{1974}).

\bibitem[{\citenamefont{Bucher and Spergel}(1999)}]{Solid}
\bibinfo{author}{\bibfnamefont{M.}~\bibnamefont{Bucher}} \bibnamefont{and}
  \bibinfo{author}{\bibfnamefont{D.~N.} \bibnamefont{Spergel}},
  \bibinfo{journal}{Phys. Rev.} \textbf{\bibinfo{volume}{D60}},
  \bibinfo{pages}{043505} (\bibinfo{year}{1999}), \eprint{astro-ph/9812022}.

\bibitem[{\citenamefont{Jones et~al.}(2002)\citenamefont{Jones, Stoica, and
  Tye}}]{Jones}
\bibinfo{author}{\bibfnamefont{N.}~\bibnamefont{Jones}},
  \bibinfo{author}{\bibfnamefont{H.}~\bibnamefont{Stoica}}, \bibnamefont{and}
  \bibinfo{author}{\bibfnamefont{S.~H.~H.} \bibnamefont{Tye}},
  \bibinfo{journal}{JHEP} \textbf{\bibinfo{volume}{07}}, \bibinfo{pages}{051}
  (\bibinfo{year}{2002}), \eprint{hep-th/0203163}.

\bibitem[{\citenamefont{Sarangi and Tye}(2002)}]{Sarangi}
\bibinfo{author}{\bibfnamefont{S.}~\bibnamefont{Sarangi}} \bibnamefont{and}
  \bibinfo{author}{\bibfnamefont{S.~H.~H.} \bibnamefont{Tye}},
  \bibinfo{journal}{Phys. Lett.} \textbf{\bibinfo{volume}{B536}},
  \bibinfo{pages}{185} (\bibinfo{year}{2002}), \eprint{hep-th/0204074}.

\bibitem[{\citenamefont{Matsuda}(2004)}]{Matsuda}
\bibinfo{author}{\bibfnamefont{T.}~\bibnamefont{Matsuda}},
  \bibinfo{journal}{JHEP} \textbf{\bibinfo{volume}{10}}, \bibinfo{pages}{042}
  (\bibinfo{year}{2004}), \eprint{hep-ph/0406064}.

\bibitem[{\citenamefont{Barnaby et~al.}(2005)\citenamefont{Barnaby, Berndsen,
  Cline, and Stoica}}]{Barnaby}
\bibinfo{author}{\bibfnamefont{N.}~\bibnamefont{Barnaby}},
  \bibinfo{author}{\bibfnamefont{A.}~\bibnamefont{Berndsen}},
  \bibinfo{author}{\bibfnamefont{J.~M.} \bibnamefont{Cline}}, \bibnamefont{and}
  \bibinfo{author}{\bibfnamefont{H.}~\bibnamefont{Stoica}},
  \bibinfo{journal}{JHEP} \textbf{\bibinfo{volume}{06}}, \bibinfo{pages}{075}
  (\bibinfo{year}{2005}), \eprint{hep-th/0412095}.

\bibitem[{\citenamefont{Oliveira et~al.}(2005)\citenamefont{Oliveira, Martins,
  and Avelino}}]{Oliveira1}
\bibinfo{author}{\bibfnamefont{J.~C. R.~E.} \bibnamefont{Oliveira}},
  \bibinfo{author}{\bibfnamefont{C.~J. A.~P.} \bibnamefont{Martins}},
  \bibnamefont{and} \bibinfo{author}{\bibfnamefont{P.~P.}
  \bibnamefont{Avelino}}, \bibinfo{journal}{Phys. Rev.}
  \textbf{\bibinfo{volume}{D71}}, \bibinfo{pages}{083509}
  (\bibinfo{year}{2005}), \eprint{hep-ph/0410356}.

\bibitem[{\citenamefont{Avelino et~al.}(2005)\citenamefont{Avelino, Oliveira,
  and Martins}}]{Oliveira2}
\bibinfo{author}{\bibfnamefont{P.~P.} \bibnamefont{Avelino}},
  \bibinfo{author}{\bibfnamefont{J.~C. R.~E.} \bibnamefont{Oliveira}},
  \bibnamefont{and} \bibinfo{author}{\bibfnamefont{C.~J. A.~P.}
  \bibnamefont{Martins}}, \bibinfo{journal}{Phys. Lett.}
  \textbf{\bibinfo{volume}{B610}}, \bibinfo{pages}{1} (\bibinfo{year}{2005}),
  \eprint{hep-th/0503226}.

\bibitem[{\citenamefont{Martins and Shellard}(1996{\natexlab{a}})}]{ms1a}
\bibinfo{author}{\bibfnamefont{C.~J. A.~P.} \bibnamefont{Martins}}
  \bibnamefont{and} \bibinfo{author}{\bibfnamefont{E.~P.~S.}
  \bibnamefont{Shellard}}, \bibinfo{journal}{Phys. Rev.}
  \textbf{\bibinfo{volume}{D53}}, \bibinfo{pages}{575}
  (\bibinfo{year}{1996}{\natexlab{a}}), \eprint{hep-ph/9507335}.

\bibitem[{\citenamefont{Martins and Shellard}(1996{\natexlab{b}})}]{ms1b}
\bibinfo{author}{\bibfnamefont{C.~J. A.~P.} \bibnamefont{Martins}}
  \bibnamefont{and} \bibinfo{author}{\bibfnamefont{E.~P.~S.}
  \bibnamefont{Shellard}}, \bibinfo{journal}{Phys. Rev.}
  \textbf{\bibinfo{volume}{D54}}, \bibinfo{pages}{2535}
  (\bibinfo{year}{1996}{\natexlab{b}}), \eprint{hep-ph/9602271}.

\bibitem[{\citenamefont{Martins and Shellard}(2002)}]{extend}
\bibinfo{author}{\bibfnamefont{C.~J. A.~P.} \bibnamefont{Martins}}
  \bibnamefont{and} \bibinfo{author}{\bibfnamefont{E.~P.~S.}
  \bibnamefont{Shellard}}, \bibinfo{journal}{Phys. Rev.}
  \textbf{\bibinfo{volume}{D65}}, \bibinfo{pages}{043514}
  (\bibinfo{year}{2002}), \eprint{hep-ph/0003298}.

\bibitem[{\citenamefont{Press et~al.}(1989)\citenamefont{Press, Ryden, and
  Spergel}}]{Press}
\bibinfo{author}{\bibfnamefont{W.~H.} \bibnamefont{Press}},
  \bibinfo{author}{\bibfnamefont{B.~S.} \bibnamefont{Ryden}}, \bibnamefont{and}
  \bibinfo{author}{\bibfnamefont{D.~N.} \bibnamefont{Spergel}},
  \bibinfo{journal}{Astrophys. J.} \textbf{\bibinfo{volume}{347}},
  \bibinfo{pages}{590} (\bibinfo{year}{1989}).

\bibitem[{\citenamefont{Martins}(2004)}]{nonint}
\bibinfo{author}{\bibfnamefont{C.~J. A.~P.} \bibnamefont{Martins}},
  \bibinfo{journal}{Phys. Rev.} \textbf{\bibinfo{volume}{D70}},
  \bibinfo{pages}{107302} (\bibinfo{year}{2004}), \eprint{hep-ph/0410326}.

\bibitem[{\citenamefont{Gregory et~al.}(1990)\citenamefont{Gregory, Haws, and
  Garfinkle}}]{ghg}
\bibinfo{author}{\bibfnamefont{R.}~\bibnamefont{Gregory}},
  \bibinfo{author}{\bibfnamefont{D.}~\bibnamefont{Haws}}, \bibnamefont{and}
  \bibinfo{author}{\bibfnamefont{D.}~\bibnamefont{Garfinkle}},
  \bibinfo{journal}{Phys. Rev.} \textbf{\bibinfo{volume}{D42}},
  \bibinfo{pages}{343} (\bibinfo{year}{1990}).

\bibitem[{\citenamefont{Kawano}(1990)}]{kawano}
\bibinfo{author}{\bibfnamefont{L.}~\bibnamefont{Kawano}},
  \bibinfo{journal}{Phys. Rev.} \textbf{\bibinfo{volume}{D41}},
  \bibinfo{pages}{1013} (\bibinfo{year}{1990}).

\bibitem[{\citenamefont{Austin et~al.}(1993)\citenamefont{Austin, Copeland, and
  Kibble}}]{ack}
\bibinfo{author}{\bibfnamefont{D.}~\bibnamefont{Austin}},
  \bibinfo{author}{\bibfnamefont{E.~J.} \bibnamefont{Copeland}},
  \bibnamefont{and} \bibinfo{author}{\bibfnamefont{T.~W.~B.}
  \bibnamefont{Kibble}}, \bibinfo{journal}{Phys. Rev.}
  \textbf{\bibinfo{volume}{D48}}, \bibinfo{pages}{5594} (\bibinfo{year}{1993}),
  \eprint{hep-ph/9307325}.

\bibitem[{\citenamefont{Hindmarsh}(2003)}]{Hindmarsh}
\bibinfo{author}{\bibfnamefont{M.}~\bibnamefont{Hindmarsh}},
  \bibinfo{journal}{Phys. Rev.} \textbf{\bibinfo{volume}{D68}},
  \bibinfo{pages}{043510} (\bibinfo{year}{2003}), \eprint{hep-ph/0207267}.

\bibitem[{\citenamefont{Moore et~al.}(2002)\citenamefont{Moore, Shellard, and
  Martins}}]{moore}
\bibinfo{author}{\bibfnamefont{J.~N.} \bibnamefont{Moore}},
  \bibinfo{author}{\bibfnamefont{E.~P.~S.} \bibnamefont{Shellard}},
  \bibnamefont{and} \bibinfo{author}{\bibfnamefont{C.~J. A.~P.}
  \bibnamefont{Martins}}, \bibinfo{journal}{Phys. Rev.}
  \textbf{\bibinfo{volume}{D65}}, \bibinfo{pages}{023503}
  (\bibinfo{year}{2002}), \eprint{hep-ph/0107171}.

\bibitem[{\citenamefont{Pen et~al.}(1997)\citenamefont{Pen, Seljak, and
  Turok}}]{pst}
\bibinfo{author}{\bibfnamefont{U.-L.} \bibnamefont{Pen}},
  \bibinfo{author}{\bibfnamefont{U.}~\bibnamefont{Seljak}}, \bibnamefont{and}
  \bibinfo{author}{\bibfnamefont{N.}~\bibnamefont{Turok}},
  \bibinfo{journal}{Phys. Rev. Lett.} \textbf{\bibinfo{volume}{79}},
  \bibinfo{pages}{1611} (\bibinfo{year}{1997}), \eprint{astro-ph/9704165}.

\bibitem[{\citenamefont{Vincent et~al.}(1998)\citenamefont{Vincent, Antunes,
  and Hindmarsh}}]{vhs2}
\bibinfo{author}{\bibfnamefont{G.}~\bibnamefont{Vincent}},
  \bibinfo{author}{\bibfnamefont{N.~D.} \bibnamefont{Antunes}},
  \bibnamefont{and}
  \bibinfo{author}{\bibfnamefont{M.}~\bibnamefont{Hindmarsh}},
  \bibinfo{journal}{Phys. Rev. Lett.} \textbf{\bibinfo{volume}{80}},
  \bibinfo{pages}{2277} (\bibinfo{year}{1998}), \eprint{hep-ph/9708427}.

\bibitem[{\citenamefont{Bennett and Bouchet}(1990)}]{bb}
\bibinfo{author}{\bibfnamefont{D.~P.} \bibnamefont{Bennett}} \bibnamefont{and}
  \bibinfo{author}{\bibfnamefont{F.~R.} \bibnamefont{Bouchet}},
  \bibinfo{journal}{Phys. Rev.} \textbf{\bibinfo{volume}{D41}},
  \bibinfo{pages}{2408} (\bibinfo{year}{1990}).

\bibitem[{\citenamefont{Allen and Shellard}(1990)}]{as}
\bibinfo{author}{\bibfnamefont{B.}~\bibnamefont{Allen}} \bibnamefont{and}
  \bibinfo{author}{\bibfnamefont{E.~P.~S.} \bibnamefont{Shellard}},
  \bibinfo{journal}{Phys. Rev. Lett.} \textbf{\bibinfo{volume}{64}},
  \bibinfo{pages}{119} (\bibinfo{year}{1990}).

\bibitem[{\citenamefont{Martins et~al.}(2004)\citenamefont{Martins, Moore, and
  Shellard}}]{prl}
\bibinfo{author}{\bibfnamefont{C.~J. A.~P.} \bibnamefont{Martins}},
  \bibinfo{author}{\bibfnamefont{J.~N.} \bibnamefont{Moore}}, \bibnamefont{and}
  \bibinfo{author}{\bibfnamefont{E.~P.~S.} \bibnamefont{Shellard}},
  \bibinfo{journal}{Phys. Rev. Lett.} \textbf{\bibinfo{volume}{92}},
  \bibinfo{pages}{251601} (\bibinfo{year}{2004}), \eprint{hep-ph/0310255}.

\bibitem[{\citenamefont{Coulson et~al.}(1996)\citenamefont{Coulson, Lalak, and
  Ovrut}}]{Coulson}
\bibinfo{author}{\bibfnamefont{D.}~\bibnamefont{Coulson}},
  \bibinfo{author}{\bibfnamefont{Z.}~\bibnamefont{Lalak}}, \bibnamefont{and}
  \bibinfo{author}{\bibfnamefont{B.~A.} \bibnamefont{Ovrut}},
  \bibinfo{journal}{Phys. Rev.} \textbf{\bibinfo{volume}{D53}},
  \bibinfo{pages}{4237} (\bibinfo{year}{1996}).

\bibitem[{\citenamefont{Larsson et~al.}(1997)\citenamefont{Larsson, Sarkar, and
  White}}]{Larsson}
\bibinfo{author}{\bibfnamefont{S.~E.} \bibnamefont{Larsson}},
  \bibinfo{author}{\bibfnamefont{S.}~\bibnamefont{Sarkar}}, \bibnamefont{and}
  \bibinfo{author}{\bibfnamefont{P.~L.} \bibnamefont{White}},
  \bibinfo{journal}{Phys. Rev.} \textbf{\bibinfo{volume}{D55}},
  \bibinfo{pages}{5129} (\bibinfo{year}{1997}), \eprint{hep-ph/9608319}.

\bibitem[{\citenamefont{Avelino and Martins}(2000)}]{Fossils}
\bibinfo{author}{\bibfnamefont{P.~P.} \bibnamefont{Avelino}} \bibnamefont{and}
  \bibinfo{author}{\bibfnamefont{C.~J. A.~P.} \bibnamefont{Martins}},
  \bibinfo{journal}{Phys. Rev.} \textbf{\bibinfo{volume}{D62}},
  \bibinfo{pages}{103510} (\bibinfo{year}{2000}),
  \eprint[http://arXiv.org/abs]{astro-ph/0003231}.

\bibitem[{\citenamefont{Avelino et~al.}(2001)\citenamefont{Avelino, Carvalho,
  Martins, and Oliveira}}]{Inhomog}
\bibinfo{author}{\bibfnamefont{P.~P.} \bibnamefont{Avelino}},
  \bibinfo{author}{\bibfnamefont{J.~P.~M.} \bibnamefont{Carvalho}},
  \bibinfo{author}{\bibfnamefont{C.~J. A.~P.} \bibnamefont{Martins}},
  \bibnamefont{and} \bibinfo{author}{\bibfnamefont{J.~C. R.~E.}
  \bibnamefont{Oliveira}}, \bibinfo{journal}{Phys. Lett.}
  \textbf{\bibinfo{volume}{B515}}, \bibinfo{pages}{148} (\bibinfo{year}{2001}),
  \eprint{astro-ph/0004227}.

\bibitem[{\citenamefont{Garagounis and Hindmarsh}(2003)}]{Garagounis}
\bibinfo{author}{\bibfnamefont{T.}~\bibnamefont{Garagounis}} \bibnamefont{and}
  \bibinfo{author}{\bibfnamefont{M.}~\bibnamefont{Hindmarsh}},
  \bibinfo{journal}{Phys. Rev.} \textbf{\bibinfo{volume}{D68}},
  \bibinfo{pages}{103506} (\bibinfo{year}{2003}), \eprint{hep-ph/0212359}.

\end{thebibliography}

\end{document}